\documentclass[aps,showpacs,twocolumn,superscriptaddress]{revtex4} 
 
\usepackage{graphicx} 
\usepackage{dcolumn} 
\usepackage{amsmath}

\newcommand{\vp}{{\vec p}}

\begin{document} 
 
\preprint{\parbox[t]{45mm}{\small ANL-PHY-10258-TH-2002\\ MPG-VT-UR~220/02}} 
 
\title{Quantum effects with an X-ray free electron laser} 
 
\author{C.D.~Roberts} 
\affiliation{Physics Division, Bldg 203, Argonne National Laboratory, Argonne 
Illinois 60439-4843} 
\affiliation{Fachbereich Physik, Universit\"at Rostock, D-18051 Rostock, 
Germany} 
 
\author{S.M.~Schmidt} 
\affiliation{Institut f\"ur Theoretische Physik, Universit\"at T\"ubingen, Auf 
der Morgenstelle 14, D-72076 T\"ubingen, Germany} 
\affiliation{Helmholtz-Gemeinschaft, Ahrstrasse 45, 
  D-53175 Bonn, Germany} 
\author{D.V.~Vinnik} 
\affiliation{Institut f\"ur Theoretische Physik, Universit\"at T\"ubingen, Auf 
der Morgenstelle 14, D-72076 T\"ubingen, Germany} 
 
\begin{abstract} 
\rule{0ex}{3ex} 
A quantum kinetic equation coupled with Maxwell's equation is used to estimate 
the laser power required at an XFEL facility to expose intrinsically quantum 
effects in the process of QED vacuum decay via spontaneous pair production. A 
$9\,$TW-peak XFEL laser with photon energy $8.3\,$keV could be sufficient to 
initiate particle accumulation and the consequent formation of a plasma of 
spontaneously produced pairs.  The evolution of the particle number in the 
plasma will exhibit non-Markovian aspects of the strong-field pair production 
process and the plasma's internal currents will generate an electric field 
whose interference with that of the laser leads to plasma oscillations. 
\end{abstract} 
 
\maketitle

X-ray free electron laser (XFEL) facilities are planned at SLAC~\cite{slac}: 
namely the Linac Coherent Light Source (LCLS), and as part of the $e^-\,e^+$ 
linear collider project (TESLA) at DESY~\cite{desy}.  They propose to provide 
narrow bandwidth, high power, short-length laser X-ray pulses, with good 
spatial coherence and tunable energy.  It is anticipated that the realisable 
values of these parameters will enable studies of completely new fields in 
X-ray science, with applications in atomic and molecular physics, plasma 
physics, and many other fields~\cite{desy}. 
 
A unique ability of these facilities is to provide very high peak power 
densities. For example, a $P=0.2\,$TW-peak laser at a wavelength of 
$\lambda=0.4\,$nm, values which are reckoned achievable with current 
technology~\cite{desy}, can conceivably produce a peak electric field strength 
\begin{equation} 
E_a = \sqrt{ \mu_0 c\, P/(\pi \lambda^2)} \approx 1.2\times 10^{16} \, {\rm 
V/m}\,. 
\end{equation} 
Boosting $P$ to $1\,$TW and reducing $\lambda$ to $0.1\,$nm, which is 
theoretically possible~\cite{Chen}, would yield an order-of-magnitude increase: 
$E_g = 1.1 \times 10^{17} \,$V/m.  Electric fields of this strength are 
sufficient for an experimental verification of the spontaneous decay of the QED 
vacuum~\cite{Fried,Ringwald,popov,bastiprl}. 
 
It is a long standing prediction that the QED vacuum is unstable in the 
presence of a strong, constant electric field, decaying via the production of 
$e^-\,e^+$ pairs~\cite{Schwinger}.  In such fields, appreciable particle 
production is certain if the strength exceeds $E_{\rm cr}:= m_e^2/e = 1.3 
\times 10^{18}\, {\rm V/m}$. (We subsequently use $\hbar = c = 1$.)  The 
proposed XFEL facilities could generate $E \approx 0.1\,E_{\rm cr}$.  (NB.\ 
Here ``constant'' means that the field must be uniform over time- and 
length-scales much greater than the electron's Compton wavelength: $1/m_e 
\approx 0.4\,$pm.) 
 
A single laser beam cannot produce pairs~\cite{Troup}.  (For a light-like 
field $F_{\mu\nu} F^{\mu\nu} = 0$ and hence the vacuum survival probability 
is equal to one.)  Nevertheless, if two or more coherent beams are crossed 
and form a standing wave at their intersection, one can hypothetically 
produce a region in which there is a strong electric field but no magnetic 
field.  The radius of this spot volume is diffraction limited to be larger 
than the laser beams' wavelength: $r_\sigma \gtrsim \lambda$, and the 
interior electric field could be approximately constant on length-scales 
approaching this magnitude.  The period of the electric field is also 
determined by $\lambda$.  Hence at an XFEL facility one might satisfy the 
length-scale uniformity conditions noted above. 
 
However, the planned laser pulse duration: $\tau_{\rm coh} \sim 100\,$fs, is 
large compared to the laser period: $\tau_\gamma \sim 1\,$as, and thus the 
time-dependence of the electric field in the standing wave may materially 
affect the pattern of observed pair production; i.e., vacuum decay might be 
expressed via time-dependent pair production.  This possibility can only be 
explored using methods of non-equilibrium quantum field 
theory~\cite{BrezinPopov} or an equivalent quantum kinetic 
theory~\cite{basti,kme,sms}.  The latter procedure was employed in 
Ref.~\cite{bastiprl}, wherein it was shown that pair production occurs in 
cycles that proceed in tune with the laser frequency.  While that does not lead 
to significant particle accumulation, the peak density of produced pairs is 
frequency independent, with the consequence that several hundred pairs could be 
produced per laser period. 
 
The proposed XFEL facilities offer a first real chance of observing the decay 
of QED's vacuum, a profound and nonperturbative quantum field theoretical 
effect.  Nevertheless, with a quantum Vlasov equation, one can ask for 
more. This equation yields the time-dependence of the single particle 
distribution function: $f(\vec{p},t):= \langle 
a^\dagger_{\vec{p}}(t)\,a_{\vec{p}}(t) \rangle$, and hence can be used to 
estimate the laser power required to achieve an accumulation of $e^-\,e^+$ 
pairs via vacuum decay.  Furthermore, in quantum field theory the particle 
production process is necessarily non-local in time; i.e., 
\textit{non-Markovian}, and dependent on the particles' statistics.  These 
features are preserved by the source term in the Vlasov equation. (The 
Schwinger source term is recovered in a carefully controlled weak-field 
limit~\cite{kme}.)  Consequently, one can identify the laser parameters 
necessary to expose negative energy elements in the particle wave packets. 
The system exhibits this core quantum field theoretical feature when the time 
between production events is commensurate with the electron's Compton 
wavelength.  Along with accumulation comes the possibility of plasma 
oscillations, generated by feedback between the laser-produced electric field 
and the field associated with the production and motion of the $e^-\,e^+$ 
pairs; and also collisions.  (These features are reviewed in 
Ref.~\cite{bastirev}.) 
 
We model an ``ideal experiment,'' and assume $r_\sigma = \lambda$ and a nonzero 
electric field constant throughout this volume while the magnetic field 
vanishes identically.  This is impossible to achieve in practice and therefore 
the field strengths actually achievable will be weaker than we suppose.  Hence 
our estimates of the laser parameters will be lower bounds. Our model is 
represented by 
\begin{equation} 
\label{laserE} {\vec E}_{\rm ext}(t) = (0,0,E_0\, \sin\Omega t\,)\, , \; \Omega = 
2\pi/\lambda\,. 
\end{equation} 
Particle production occurs throughout the coherence spike length, $\tau_{\rm 
coh}$. However, collisions and radiation may become important as time 
progresses and therefore, to simplify our analysis, we focus on the first 
$100$ laser periods; i.e., the first $\sim 0.1$\% of the laser pulse. 
 
In spatially homogeneous fields the kinetic equation has the form 
\begin{equation} 
\frac{d\,f(\vp,t)}{dt}=S(\vp,t)+C(\vp,t)\,, 
\end{equation} 
where $S(\vp,t)$ is a source term describing particle production and 
$C(\vp,t)$ is a collision term, which properly includes annihilation.  The 
form is intuitive: the rate of change in the single particle distribution 
function, $f(\vec{p},t)$, is determined by competition between particle 
production, and collisions and annihilation.  The source term can be 
calculated directly using quantum mean field theory~\cite{basti,kme}.  The 
collision term, however, is more complicated and not directly accessible in 
the mean field approximation. It only becomes important, though, in dense 
plasmas and because we deliberately avoid that situation, we proceed by 
setting $C\equiv 0$. 
 
The distribution of particles is described by $f(\vec{p},t)$.  These 
particles are accelerated by $E_{\rm ext}(t)$, and the associated currents 
generate an opposing electric field, $E_{\rm int}(t)$.  It is the sum: 
$E(t)=E_{\rm ext}(t)+E_{\rm int}(t)$, which is finally responsible for 
particle production. Thus, to properly represent the system's evolution, one 
must solve a Vlasov equation coupled with Maxwell's equation~\cite{epjc}: 
\begin{eqnarray} 
\label{KE} 
\frac{df(\vec{p},t)}{dt} &=& \frac{eE(t)\varepsilon_\perp^2}{2\omega^2 (\vp,t)} 
\int_{0}^t dt'\frac{eE(t')\big[1-2f(\vp,t')\big]}{\omega^2 (\vp,t^\prime)} 
\nonumber \\ 
&& \qquad\qquad \times \cos\left[ 2\int_{t'}^t d\tau ~\omega (\vp,\tau) \right] 
, 
\\ 
{\dot E}_{int}(t) &=& -4 e \int\!\frac{d^3p}{(2\pi)^3}\,\frac{1}{\omega(\vp,t)} 
\Biggl[ p_\| f(\vp,t) 
\nonumber \\ 
&& + \frac{\omega^2(\vp,t)}{eE(t)}\frac{d\, f(\vp,t)}{d\,t} - \,\frac{e \dot 
E(t)\,\varepsilon_\perp ^2}{8\, \omega^4(\vp,t)} \Biggr] , \label{ME} 
\end{eqnarray} 
with: $\vp=(\vp_{\perp} ,p_3)$; $\varepsilon_{\perp}^2=m_e^2+p_{\perp}^2$; 
$\omega^2(\vp,t)=\varepsilon_{\perp}^2+p_\|^2(t)$, $p_\|(t)=p_3 - e\,A(t)$. 
 
Quantum statistics affect the production rate through the term \mbox{``$[1 - 
2 \,f]$''} in Eq.~(\ref{KE}), which ensures that no momentum state has more 
than one spin-up and one spin-down fermion.  In addition, both this factor 
and the ``$\cos$'' term introduce non-Markovian character to the system: the 
first couples in the history of the distribution function's time evolution; 
the second, that of the field. 
 
There are two control parameters in Eq.~(\ref{KE}): the laser field strength, 
$E_0$, and the wavelength, $\lambda$.  We fix $\lambda = 0.15\,$nm, which is 
achievable at the proposed XFELs.  (NB.\ By assumption, the volume in which 
particles are produced increases with $\lambda^3$ whereas the field strength 
decreases with $1/\lambda$: there is merit in optimising $\lambda$.) Our 
study will expose additional phenomena that become observable with increasing 
$E_0$. 
 
\begin{figure}[t] 
\centerline{\includegraphics[height=5.7cm]{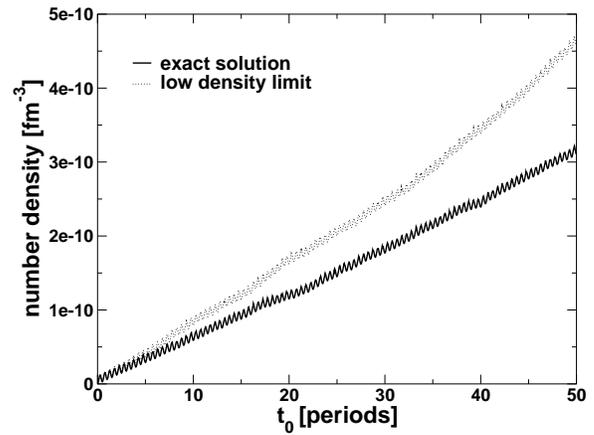}} \caption{\label{fig1} 
Number density calculated with $E_0=0.5\,E_{cr}$.  Solid line: solution of 
Eqs.~(\protect\ref{KE}), (\protect\ref{ME}); dotted line: solution obtained 
using a low density approximation to $S(\vp,t)$, Eq.~(\protect\ref{ld}).  The 
oscillations are tied to the laser frequency, see Eq.~(\protect\ref{ntfit}).} 
\end{figure} 
 
The particle number density: 
\begin{equation} 
n(t) = 2 \int \frac{d^3 p}{(2\pi)^3}\, f(\vec{p},t)\,, 
\end{equation} 
is plotted in Fig.~\ref{fig1} for $E=0.5\,E_{cr}$.  Our results for 
$T=t/\lambda \lesssim 100$ are accurately fitted by 
\begin{eqnarray} 
\label{ntfit} 
n(T;E_0) & = & a_0(E_0) \, \sin^2 2\pi T\, + \rho(T,E_0)\,T \,, \\ 
\label{kappadef} 
\rho(T,E_0) & = & \rho(E_0) + \rho^\prime(E_0)\, T \,. 
\end{eqnarray} 
We can therefore use $\rho(T,E_0)$ to quantify the rate of particle 
accumulation.  This rate is very small for $E_0 \lesssim 0.2\,E_{\rm 
cr}$~\cite{bastiprl} and, while noise in $n(t)$ prevents a reliable numerical 
determination of $\rho$, $\rho^\prime$, it is nevertheless clear that in this 
case lengthening $\tau_{\rm coh}$ will not materially increase the number of 
particles produced. 
 
It is apparent from Fig.~\ref{fig1}, however, that the situation is very 
different for $E_0= 0.5\,E_{\rm cr}$.  The solid curve is described by 
Eq.~(\ref{ntfit}), with $a_0= 1.2 \times 10^{-11}\,$fm$^{-3}$, $\rho= 5.4 
\times 10^{-12}\,$fm$^{-3}\,$period$^{-1}\!$, $\rho^\prime/\rho= 
0.0033/$period.  Clearly, the accumulation rate is approximately constant. 
 
\begin{figure}[t] 
\centerline{\includegraphics[height=5.8cm]{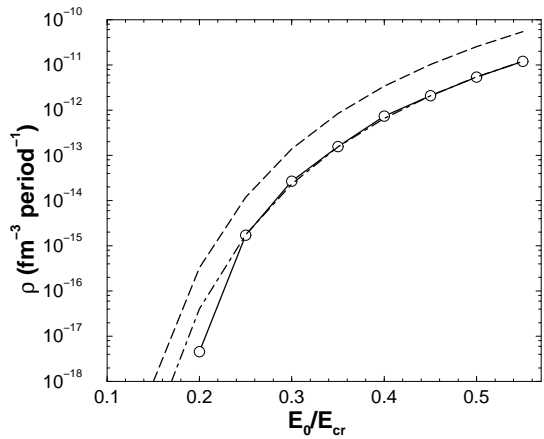}}\vspace*{-0.7ex} 
 
\caption{\label{fig2} Solid line: accumulation rate, $\rho(0,E_0)$, 
Eq.~(\protect\ref{kappadef}); dashed line: Schwinger rate, 
Eq.~(\protect\ref{schwingerrate}); dot-dashed line: 
Eq.~(\protect\ref{schwingerrate}) with $a_S\to a = 0.305$, $b_S\to b=1.06$.} 
\end{figure}

Solving the coupled system of Eqs.~(\ref{KE}), (\ref{ME}) numerically is 
straightforward but time consuming.  We have repeated the procedure for a 
range of values of $E_0$ and plotted the calculated values of $\rho(E_0)$ in 
Fig.~\ref{fig2}.  These results can be used to obtain a rough estimate of the 
accumulated number density via: $n(T,E_0) \approx \rho(E_0)\, T$. 
We also plot: 
\begin{equation} 
\label{schwingerrate} \rho_{\rm S} = a_S\,\frac{m_e^4 \lambda}{4\pi^3} 
\left[\frac{E_0}{E_{\rm cr}}\right]^2 \,{\rm e}^{-b_S\,\pi E_{\rm cr}/E_0},\; 
a_S=1,\, b_S=1\,; 
\end{equation} 
i.e., the Schwinger rate.  As can be anticipated from the energy budget, the 
constant-field Schwinger rate bounds our results from above.  This comparison 
provides a context for the estimates in Ref.~\cite{Ringwald}. 
 
\begin{figure}[t] 
\centerline{\includegraphics[height=5.7cm]{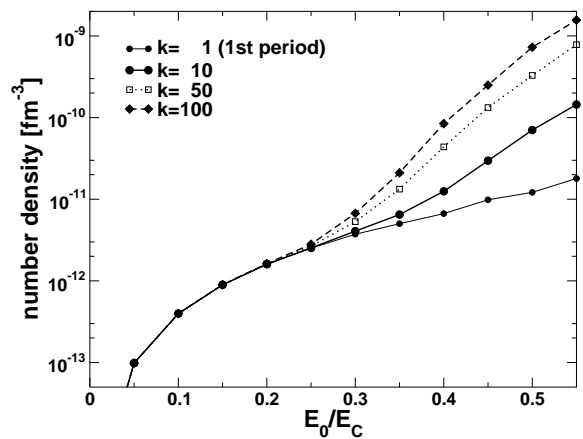}} \caption{\label{fig3} 
Peak particle number density versus laser field strength. There is a striking 
qualitative change at $E_0 \approx 0.25\,E_{\rm cr}$, which marks the onset of 
particle accumulation.} 
\end{figure}

In Fig.~\ref{fig3} we plot the \textit{peak} particle number density, 
$n(t_k^>)$, where $t_k^>= (4 k -3) \lambda/4$ is the time at which $E(t)$ is 
maximal during laser period no.\ $k$.  The figure displays a qualitative change 
in the rate of particle production with increasing laser field strength. 
[$n(k \lambda) \approx n(t_k^>)$ if and only if there is significant particle 
accumulation, otherwise $n(k \lambda)\ll n(t_k^>)$.] 
 
That change is emphasised by Table~\ref{table1}, which reports 
\begin{equation} 
\label{Npeak} N(t_k^>,E_0) := \sigma \, n(t_k^>,E_0)\,; 
\end{equation} 
i.e.,  the peak particle number. 
\begin{table}[b] 
\caption{\label{table1} Peak particle number, Eq.~(\protect\ref{Npeak}), at 
selected values of the laser field strength, $E_0$ in 
Eq.~(\protect\ref{laserE}), with $k=50$.} 
\begin{center} 
\begin{tabular}{c|c|c|c|c|c} 
 $\displaystyle E_0/E_{cr}$ &0.1 &0.2 & 0.3&0.4&0.5\\\hline 
 $\displaystyle N_\sigma\, (\times 10^3)$& ~1.3& ~5.4& ~17.9& ~148&~1110 
  \rule{0em}{3ex} 
\end{tabular} 
\end{center} 
\end{table} 
For $E_0 \lesssim 0.25 \,E_{\rm cr}$ there is no significant accumulation of 
particles: $N(t_0^>,E_0) \approx N(t_{10}^>,E_0)$, etc., just as observed in 
Ref.~\cite{bastiprl}.  However, for $E_0 > 0.25 \,E_{\rm cr}$ there are more 
particle pairs after each successive laser period; e.g., $E_0 = 0.35\,E_{\rm 
cr}$ brings an order of magnitude increase in $N$ over the first $100$ laser 
periods.  At such values of $E_0$ one is in a domain where numerous electrons 
and positrons produced in a single period are accelerated to relative 
longitudinal momenta that are sufficient to materially inhibit annihilation. 
This ensures that many pairs remain when the next burst of production occurs. 
Consequently $N(t_k^>,E_0)$ grows considerably with increasing $k$. (NB.\ 
Accumulation means collisions can become important and should be included in 
the kinetic equation if quantitatively reliable results are required.  As we 
neglect these effects we do not follow the system's evolution into the high 
density regime.) 
 
In the low density limit: $f(\vec{p},t)\ll 1$, the distribution function 
disappears from the right hand side of Eq.~(\ref{KE}), which then yields an 
algebraic solution: 
\begin{eqnarray} 
f^{\rm ld}(\vp,t) &=& \int_{0}^t d\tau \frac{eE(\tau) 
\varepsilon_\perp^2}{2\omega^2 (\vp,t)} \int_{0}^\tau dt'\frac{eE(t')}{\omega^2 
(\vp,t')} 
\nonumber \\ 
&& \qquad \times  \cos\left[ 2\int_{t'}^t d\tau' ~\omega (\vp,\tau') \right] . 
\label{ld} 
\end{eqnarray} 
We anticipated this in Fig.~\ref{fig1}, which contrasts the low-density 
result with the complete solution for $E_0 = 0.5\,E_{\rm cr}$, which we now 
know generates significant particle accumulation.  The comparison emphasises 
that, with particle accumulation, quantum statistical effects in the source 
term are \textit{observable}. These effects are essentially non-Markovian. 
The difference grows with increasing field strength because Pauli blocking 
becomes more important with increasing fermion density. 
 
We now return to Eq.~(\ref{ME}); i.e., Maxwell's equation: $\dot E = - j$. 
The right hand side has two components: the first term, proportional to 
$f(\vec{p},t)$, is a conduction current tied to the particles' motion; and 
the last two terms express a polarisation current that is tied to the pair 
production rate~\cite{bloch}.  While these terms remain small the total 
electric field is determined solely by the applied laser beams.  However, in 
the domain of particle accumulation the internal currents can modify the 
electric field in the system, so that the coupling between Eqs.~(\ref{KE}) 
and (\ref{ME}) becomes important.  This marks the onset of field-current 
feedback, which can lead to plasma oscillations. 
 
\begin{figure}[t] 
\centerline{\includegraphics[height=5.7cm]{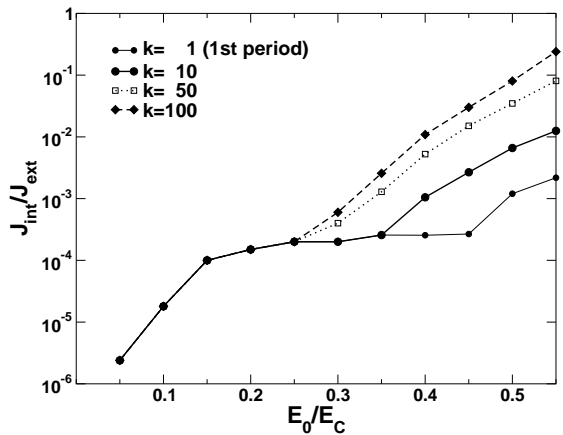}} \caption{\label{fig4} 
Ratio of the peak internal and external currents, plotted as function of 
$E_0/E_{cr}$.  Field-current feedback becomes important for $E_0 \gtrsim 
0.25\,E_{cr}$.} 
\end{figure}

Figure~\ref{fig4} depicts the ratio of the internal and external currents. 
Evidently, with the field strengths realisable at proposed XFELs, the internal 
current is negligible.  However, with the onset of particle accumulation the 
ratio grows with the number of elapsed laser periods and, since there are 
roughly $10^5$ periods in a given pulse, feedback will necessarily come to 
affect the plasma's behaviour at some stage in its evolution. Notably, for $E_0 
= 0.5\,E_{\rm cr}$ that happens very early; i.e., after only $0.1\,$\% of the 
lifetime of the pulse, and feedback will have observable consequences.  (NB.\ 
Our full calculations incorporate this effect but the result is a signal that 
less complete estimates will become more unreliable in the domain of particle 
accumulation.)  A proper analysis of the observable effects of feedback must 
also include an accurate description of collisions, which can act to damp 
plasma oscillations~\cite{epjc}. 
 
We have explored the possibility of $e^-\,e^+$ pair production using XFEL beams 
as a parameter-free application of nonequilibrium quantum mean-field theory. An 
idealised model of the laser electric field was used as input to a quantum 
Vlasov equation coupled with Maxwell's equation.  Decay of the QED vacuum is 
likely to be observed at proposed XFEL facilities.  However, we showed that 
additional phenomena become observable if the peak power is increased by a 
factor of $\sim 9$.  Then the particles produced from the QED vacuum accumulate 
to form a plasma, which is not achieved at the DESY and SLAC facilities. That 
plasma's behaviour will expose essential quantum field theoretical aspects of 
the pair production process, which are absent from the Schwinger formula. While 
this power is insufficient to expose negative energy components in the electron 
wave packets, it can reveal the non-Markovian nature of pair production; i.e., 
the feature that the appearance of a new pair is not independent of the 
production of the preceding pairs nor the evolution of the plasma before this 
new pair is produced.  With this peak power the accumulated charged pairs are 
also likely to generate internal currents that interfere with the applied laser 
field so that the plasma exhibits charge-density fluctuations; i.e., plasma 
oscillations.  An accurate prediction of this effect's observable consequences 
requires a realistic collision term in addition to the proper source.

%
\noindent We thank R.~Alkofer and A.~Ringwald for dis\-cussions.  This work 
was supported by the Deutsche For\-schungs\-ge\-mein\-schaft, under contract 
nos.\ Ro~1146/3-1 and SCHM~1342/3-1; the US Department of Energy, Nuclear 
Physics Division, under contract no.~\mbox{W-31-109-ENG-38}; the US National 
Science Foundation under grant no.\ INT-0129236; and benefited from the 
resources of the US National Energy Research Scientific Computing Center. 
 

\end{document}